# Instability of cumulation in converging cylindrical shock wave

Sergey G. Chefranov

Physics Department, Technion-Israel Insitute of Technology, Haifa 32000, Israel

csergei@technion.ac.il

## Abstract

The conditions of linear instability for a converging cylindrical shock wave in an arbitrary inviscid medium are obtained. The initial continuous cylindrical symmetry of the shock wave front is exchanged on a discrete symmetry which is determined by the most unstable small azimuthal dimensionless wave numbers $0 < k < k_{th} < 1$ of corrugation perturbations. Due to the long azimuthal wavelengths ( $\lambda = 2\pi R_{s0} / k$ , $R_{s0}$ - the radius of the shock wave) of perturbations the shape of the resulting shock wave front is not change significantly, but the corresponding restriction of the internal energy cumulation can be caused by intensification of the rotation of the medium behind the front. The instability and the restriction of cumulation are also possible in the case of the exponential rapid growth of the one-dimensional perturbations with $k = 0$ , when the shape of the shock front is not changed at all. The correspondence of present theory to the experimental and simulation data on underwater electrical explosion is considered.

## Introduction

The stability of converging spherical and cylindrical shock waves is of fundamental and applied importance for high-pressures physics and astrophysics in connection with the problem of energy accumulation [1]-[15]. It is assumed that converging shock waves are unstable with respect to extremely small perturbations, but due to the slow power-law (not exponential) growth of perturbations over time, such instability cannot prevent unlimited accumulation of pressure and temperature during implosion [10], [11]. Indeed, the linear instability of a converging spherical shock wave in an ideal gas with an adiabatic exponent $\gamma = 5/3$ and $\gamma = 3$ is established in [7]. In this case, the increase in the amplitude of small perturbations is inversely proportional to the radius of the converging shock wave $R^{-\lambda}(t)$, where the small index $0 < \lambda < 1$ increases only slightly with an increase in the adiabatic index $\gamma$ . Conclusions about the instability of a converging spherical shock wave in an ideal gas and in a Van der Waals gas were also obtained in [8], [13] and [11] respectively. As in [7], the increase in time of the amplitude of the radial disturbance of the shock wave front in [13] (where $\gamma = 5/3$ and $\gamma = 7/5$ ) is also due to a decrease in time of the shock wave radius. Moreover, in [9], based on the consideration of finite-amplitude perturbations for a converging cylindrical shock wave, the nonlinear stability of the shape of the shock wave front is established, despite the instability in the linear approximation obtained in [3].

Indeed, the preservation of the shape of the front of a converging cylindrical shock wave is observed [12] even under large compressions. This is usually considered as a manifestation of its stability mechanism due to an increase in the velocity of the shock wave front with a decrease in the local radius of curvature of this front [6], [8]-[10]. Although on the other hand there are also examples of the implementation of geometric instability for converging cylindrical shock [14] (see Fig.2 in [14]). In this regard, in [6], the problem of the stability of the front of converging spherical and cylindrical shock waves in an ideal gas with an arbitrary adiabatic index is reduced to determining the maximum permissible deviation of the shape of the shock wave front from the initial symmetric state, at which the specified mechanism for restoring the symmetry of the front is still effective. At the same time, in [6] it is pointed out that it is impossible to solve such a problem within the framework of the standard linear stability theory with respect to extremely small perturbations in amplitude. Indeed, as shown in [6], only for the threshold finite amplitude of perturbations determine the stability limits of converging spherical and cylindrical shock waves for different Mach numbers.

On the other hand, in [5] an example of limiting the cumulation process is considered when a thin cylindrical shell of an ideal liquid moves in the radial direction to the axis and initially rotates weakly around this axis. Indeed, as the shell converges, it thickens, the rotation of the inner layers increases, and due to the centrifugal force, the shell does not reach the axis, and its expansion begins [5]. Even with a weak initial excitation of rotation around the axis of symmetry, this degree of freedom gradually captures all the energy of the initial basic radial motion. In the given example, a small tangential perturbation, although it eliminates unlimited cumulation, but the smaller this perturbation, the greater cumulation is possible for the energy density of purely rotational motion. In this regard, in [5], a statement is made about the impossibility of a real limitation of cumulation due to extremely small perturbations.

At the same time, as with the weak power-law growth of perturbations in [7], [13], in [5], the intensification of the initially small rotation is due only to the process of decreasing the radius of the cylindrical shell itself and is determined only from the law of conservation of the angular momentum. Therefore, for relatively small time intervals, at which the radius of the shell can be considered practically unchanged, the value of the initial small rotational perturbation also remains practically constant.

Thus, it would seem that it should follow from the currently available representations given above that extremely small initial perturbations of the symmetry of both the converging shock wave front itself and the symmetry of the velocity field of the medium behind the shock wave front cannot lead to a limitation of the accumulation of the internal energy density in the implosion process.

In the present paper, however, it is shown that this representation needs to be refined, since the possibility of a rapid exponential growth of extremely small perturbations of the tangential component of the velocity field of the medium behind the shock wave front is established. So, with a sufficiently large index of exponential growth in time, such perturbations can already reach finite values during the time intervals at which the radius of the shock wave does not change noticeably [15]. As a result, due to the appearance of rotation with a finite speed, as in the example [5], the process of unlimited accumulation of the internal energy density becomes unstable due to the transfer of energy into the energy of the rotational motion of the medium behind the shock wave front. However, in contrast to [5], there is no need to consider the limit, when the radius of the shock wave tends to zero.

The instability conditions obtained in [15] for a converging cylindrical shock wave, however, relate only to the case when the mechanism of this instability is associated with viscosity of the medium. Therefore, it is of interest to determine the instability conditions of a converging cylindrical shock wave in an arbitrary medium for cases when the viscosity effect can be ignored. We also note that the problem of linear stability of converging shock waves for the case of an arbitrary medium has so far been considered only in [15], where the account of the arbitrariness of the medium is carried out by analogy to the Dyakov theory, developed only for plane shock waves [16], [18], [19].

In this paper, based on the theory [15], the mechanism of instability of the process of unlimited accumulation of internal energy behind the front of a converging shock wave is considered, which is caused by the development of small perturbations of the tangential component of the velocity field. The resulting spiral, swirling motion of the medium behind the front of the converging cylindrical shock wave can be axisymmetric or almost axisymmetric and do not lead to a noticeable distortion of the initial shape of the front. A similar process of formation of swirling motion of the medium is observed in many natural and technical systems (for example, in tornadoes and dust devils), which is due to the energy optimality of swirling flows [17].

Based on the dispersion equations obtained in [15] for extremely small perturbations, the instability condition of a converging cylindrical shock wave is established for the case of zero viscosity. It is shown that the instability of the front of a cylindrical shock wave is possible not only with respect to two-dimensional (corrugating), but also with respect to one-dimensional extremely small perturbations that do not distort the shape of the front at all. Note that in [16] it is indicated that there is a need for a characteristic parameter in the system that has the dimension of the inverse time for the possibility of realizing the instability of the shock wave front relative to one-dimensional perturbations. In the absence of such a parameter, as is known, the front of a plane shock wave in an ideal inviscid medium is stable with respect to one-dimensional perturbations [16], [18], [19].

When taking into account the viscosity, as shown in [20], [21], such a parameter appears and it is equal to the ratio of the square of the local speed of sound to the coefficient of kinematic viscosity $c^2/\nu$. Accordingly, in [21] the condition of instability of a strong plane shock wave with respect to one-dimensional perturbations is established, although for a weak plane shock wave stability is maintained with respect to such perturbations [20]. In [15], the instability of a strong cylindrical shock wave to one-dimensional perturbations, as well as to two-dimensional corrugation perturbations, was similarly established. In this case, the instability is due to the well-known destabilizing effect of viscous dissipation, similar to the phenomenon of instability of Tolmin-Schlichting waves in the boundary layer [16], [22].

In this paper, the conditions for the instability of a strong cylindrical shock wave with respect to one-dimensional perturbations with zero azimuthal wave number $k = 0$ are obtained even at zero viscosity. This possibility is due to the presence of a characteristic parameter having the dimension of the inverse time and equal to $c / R_{s0}$. Here $R_{s0}$ - is the undisturbed value of the radius of the front of the cylindrical shock wave, which will be assumed to be a constant value for the considered sufficiently small time intervals.

Note that in the traditional formulation of the linear instability theory, small perturbations in a cylindrical coordinate system are characterized only by integer values of the azimuthal wave number $k = 0,1,2,...$ , which is due to the commonly used assumption that the amplitude of perturbations are unambiguously determined depending on the angular variable [16]. However, when considering perturbations of a random nature, as is in the most cases, this assumption can be abandoned and arbitrary real values of the azimuthal wave number can be used, and not only its integer values. In this case, the perturbations are described by a multi-leaf function, on each sheet of which (determined by changing the angular variable by angles that are multiples of 360 degrees) the corresponding realization of a small perturbation is already uniquely determined. In fact, this approach extends the usual use of the representation of perturbations in the form of a complex valued functions (see [16]) to functions that are not only complex valued, but also multi-valued. In this paper, it is shown that when considering two-dimensional corrugation perturbations, the instability of a converging cylindrical shock wave is established for two-dimensional perturbations with small non-integer wave numbers $0 < k < 1$ that do not exceed a certain threshold value. In this case, the realization of instability leads to an exchange of the initial continuous symmetry of the shape of the front on a discrete symmetry, which gives invariance of the resulting shock front structure with respect to turns on angles that are multiples of the value $2\pi / k$ radian. For example, at the considered threshold value $k = 0.25$, these angles are multiples of $8\pi$ radians, which provides only to a very small difference in the shape of the shock wave front from the original cylindrical symmetric shape. This conclusion corresponds to the data of observations and the corresponding simulation given in [12], where a converging cylindrical shock wave obtained during an underwater electric explosion of a wire ensemble was studied. An interpretation of the conclusion of [12] on the observed limitation of the cumulation of the internal energy density due to perturbations with a large azimuthal wavelength is proposed. It is shown that this effect is associated with an exponentially rapid increase in such perturbations over a time insufficient for a noticeable change in the radius of the shock wave. Earlier, on the contrary, the effect obtained in [12] was considered as the result of a relatively long compression time, when the radius of the shock wave changes many times, which allows the perturbations to have time to grow, even with a weak power law of increasing perturbations.

In the first section, a mathematical statement of the problem of instability of a strong converging cylindrical shock wave in an arbitrary inviscid medium with respect to two-dimensional extremely small corrugation perturbations is formulated. In the second section, the case of one-dimensional perturbations is considered. In the third section, a comparison with known experimental data is made about the threshold occurrence of instability of a converging shock wave in terms of the Mach number. The interpretation of the limitation of the cumulation of the internal energy density noted in [12], associated with perturbations with a large azimuthal wavelength, is given.

## 1. Two-dimensional (corrugation) long wave instability

Let us consider a converging cylindrical shock wave (SW) of arbitrary intensity, propagating in the directions perpendicular to the $z$ axis in the cylindrical variables $(z, r, \varphi)$ according to the theory [15].

The radial velocity of the converging shock front is $D < 0$ and $U < 0$ is the radial velocity of the medium behind the shock wave front. For simplicity, let us consider the case when it is assumed that the SW front is uniform in the direction of the $z$ axis and there are no perturbations of the velocity field component along this axis. This case allows us to simulate a converging cylindrical shock wave arising from an explosion of a system consisting of a finite number of long wires bounding a cylindrical region with an axis coinciding with the z −axis [12]. In this case, perturbations in the azimuthal direction can be caused by the finite distance between the wires, which determines the wavelength of the corresponding perturbation. Since the wires are assumed to be uniform along the length, this corresponds to the assumption made above that there are no disturbances along the z −axis. Also let us consider the quasi stationary limit when $D \approx const; U \approx const$ and it is possible to neglect the piston influence on the shock front propagation and its stability.

In this limit the equation for the perturbed cylindrical surface of the shock

$R_s = R_{s0}(\tau = \varepsilon t) + r_{sa}(\varphi, t); \varepsilon << 1, \tau \approx const$ depends only on time and the azimuthal coordinate $\varphi$ in the form (see Fig.1):

$$r_{sa} = g(\varphi, t) \qquad (1)$$

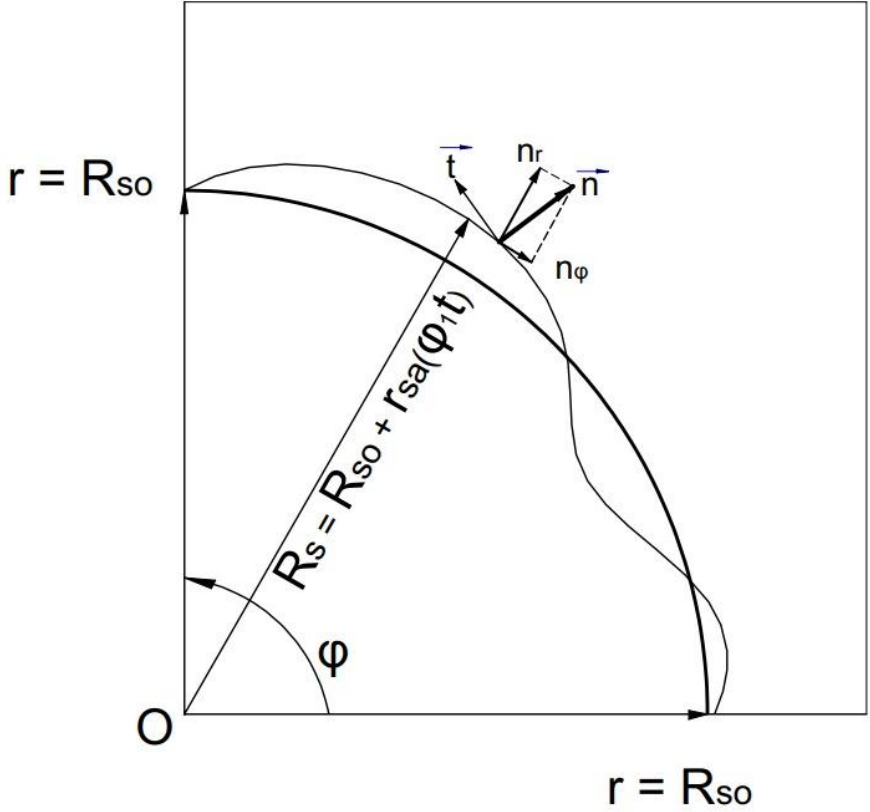

Fig.1

Caption to Fig.1: An image of a fragment of the front of a converging cylindrical shock wave is given. The value of azimuthal wave number here is equal to the value near $k = 4$. The unperturbed front is characterized by the magnitude of the shock wave radius $R_{s0}$. The resulting shape of the front after the perturbation $r_{sa}(\varphi, t)$, is applied, as well as the normal vector $\vec{n} = (n_r; n_\varphi)$ to the perturbed front surface and the tangent vector $\vec{t} = (t_r; t_\varphi)$ are given. In the drawing, the dot $O$ indicates the place to which the shock wave is converged. Reproduced with permission from Chefranov, Phys. Fluids, **32,** 114103 (2020) Copyright 2020 AIP Publishing LLC (Fig. 1 in Ref. [15]).

In (1), the value $r_{sa}$ characterizes a small additional radial deviation of the surface of the shock wave front from the unperturbed radius of the shock wave $R_{s0}$ when $r_{sa} << R_{s0} \approx const$.

Equations for small perturbations of the velocity and pressure fields behind the front of the shock wave in the linear approximation have the form in the reference of frame where shock is at rest ( $w = U - D \approx const$ ), when they are represented in cylindrical coordinates [15]:

$$\frac{\partial V_{1r}}{\partial t} + w\frac{\partial V_{1r}}{\partial r} = -\frac{1}{\rho}\frac{\partial p_1}{\partial r};$$

$$\frac{\partial V_{1\varphi}}{\partial t} + w(\frac{\partial V_{1\varphi}}{\partial r} + \frac{V_{1\varphi}}{r}) = -\frac{1}{r\rho}\frac{\partial p_1}{\partial \varphi};$$

$$\frac{\partial p_1}{\partial t} + w\frac{\partial p_1}{\partial r} + c^2\rho(\frac{1}{r}\frac{\partial r V_{1r}}{\partial r} + \frac{1}{r}\frac{\partial V_{1\varphi}}{\partial \varphi}) = 0; \qquad (2)$$

$$div\vec{V}_1 = \frac{1}{r}\frac{\partial(rV_{1r})}{\partial r} + \frac{1}{r}\frac{\partial V_{1\varphi}}{\partial \varphi}$$

In (2) the perturbation fields are denoted by the subscript 1; $c^2 = (\partial p / \partial \rho)_s$ is the square of the speed of sound in the region behind the shock wave front where the Mach number $M = |w|/c < 1$.

System (2) must be considered together with the boundary conditions on the perturbation (of near stationary cylindrical shock surface with $R_{s0} \approx const$; ) surface which is defined by function (1). It is possible to neglect the influence of viscosity on the boundary conditions in the limit of small viscosity (or large Reynolds numbers). Then, for the tangential and normal unit vectors to this surface, we have (see also [15], [18], [19]):

$$\vec{t} = (t_r, t_\varphi) = (-g_\varphi / R_{s0}; -1) / \sqrt{1 + g_\varphi^2 / R_{s0}^2};$$

$$\vec{n} = (n_r, n_\varphi) = (1; -g_\varphi / R_{s0}) / \sqrt{1 + g_\varphi^2 / R_{s0}^2}; \qquad (3)$$

$$g_\varphi = \frac{\partial g(\varphi, t)}{\partial \varphi}$$

In the linear approximation, one can neglect nonlinear terms in (3). From the condition of continuity of the tangential component of the velocity field on the perturbed front of a shock wave, it follows the boundary condition in a linear approximation in terms of the perturbation amplitudes:

$$V_{1\varphi} = g_\varphi(w_0 - w) / R_{s0} \qquad (4)$$

Similarly, the boundary condition for the normal component of the velocity field leads to the equation:

$$V_{1r} = \frac{(w - w_0)}{2} \left( \frac{p_1}{p - p_0} + \frac{\rho_1}{\rho^2 (1/\rho_0 - 1/\rho)} \right) \tag{5}$$

Let us determine the relationship of perturbation of density and pressure in (5). In this connection, in [15], [16], [18],-[19], it is proposed to use of the relation between density and pressure perturbations on the Hugoniot shock curve in the form of $p_1 = (\frac{dp}{d(1/\rho)})_H (-\frac{\rho_1}{\rho^2})$ or:

$$\rho_1 = -p_1 h \frac{\delta^2}{w_0^2};$$
$$h = j^2 \left( \frac{d(1/\rho)}{dp} \right)_H ; \, j^2 = \frac{p - p_0}{1/\rho_0 - 1/\rho} \tag{6}$$

To find the equation that determines function (1), we use the equality determining the perturbation of the velocity of the shock wave in the form $D_1 = \partial g / \partial t$ and after dropping the quadratic terms in the perturbation of density and pressure gives the equation:

$$\frac{\partial g}{\partial t} = -\frac{w_0}{2} \left( \frac{p_1}{p - p_0} - \frac{\rho_1}{\rho^2 (1/\rho_0 - 1/\rho)} \right) \tag{7}$$

As in (5), in (7), the relation between density and pressure perturbations is given by relation (6).

Let us look for a solution of equations (2) with boundary conditions (4), (5) and relations (6), (7) as

$$g(\varphi, t) = \bar{g} \exp(i(k\varphi - \omega t));$$
$$(V_{1r}, V_{1\varphi}, p_1) = (\bar{V}_r, \bar{V}_\varphi, \bar{p}) K_1(lr) \exp(i(k\varphi - \omega t)) \tag{8}$$

In (8), $K_1(lr)$ is the MacDonald function (modified Bessel's function of second kind) of first-order that gives zero boundary conditions for perturbations at infinity $r \to \infty$ and $l$ is always positive ($l > 0$)and it is assumed, as usual, that for all physical quantities in (8), the real part of the representation used is considered.

After substitution of (10) in the resulting system of equations, from (5) - (6) we obtain the following dispersion equation:

$$\omega + Mcl_1(1 - \frac{(1-h)l_2}{2M^2 l_1}) = \frac{(1+h)k^2 c^2 \delta}{2R_{s0}^2 \omega} \qquad (9)$$

$$l_1 \equiv il(\frac{K_0(lR_{s0})}{K_1(lR_{s0})} + \frac{1}{lR_{s0}}); l_2 \equiv il\frac{K_0(lR_{s0})}{K_1(lR_{s0})}$$

In (9) $K_1$ and $K_0$ - modified Bessel functions of the first and zero order, respectively.

Dispersion equation (9) must be considered together with the dispersion equation which is defined after substitution of (8) in (2) and integration of all equations $\int_0^\infty dr r^3$ (when relations

$\int_0^\infty dr r^{\mu-1} K_\nu(lr) = \frac{2^{\mu-2}}{l^\mu} \Gamma(\frac{\mu+\nu}{2}) \Gamma(\frac{\mu-\nu}{2})$ are taken into account, where $\Gamma$ is the gamma function):

As a result, we obtain a dispersion equation that has the form (see the dispersion equation (16) in [15] in the limit of zero shear and volume viscosity):

$$\left(\Omega + 2Mcl\right)\left(\Omega^2 - 24c^2 l^2\right) + 4k^2 c^2 l^2 \Omega = 0;$$
$$\Omega = i\omega \frac{3\pi}{2} - 6Mcl \qquad (10)$$

Taking into account the representation $\omega = i\omega_1$, this equation reduces to a cubic equation having the form:

$$y^3 + py + q = 0;$$
$$y = \frac{8}{9} + \frac{\pi\omega_1}{4Mcl}; p = -\frac{1}{27}\left(1 + \frac{3(k^2+6)}{M^2}\right); q = -\frac{1}{81}\left(\frac{2}{9} + \frac{k^2-12}{M^2}\right) \qquad (11)$$

The solutions of equation (11) according to the Cardano formula have the form:

$$y = y_1 = \alpha^+ + \alpha^-;$$

$$y = y_\pm = -\frac{\alpha^+ + \alpha^-}{2} \pm i\frac{\sqrt{3}}{2}\left(\alpha^+ - \alpha^-\right) \tag{12}$$

$$\alpha^\pm = \left(-\frac{q}{2} \pm \sqrt{Q}\right)^{1/3}; Q = \left(\frac{p}{3}\right)^3 + \left(\frac{q}{2}\right)^2$$

In (12) for any $k$ and $M$ the value $Q(k;M) < 0$, that corresponds to three different real solutions. Therefore, according to (12), these solutions have the form:

$$y = y_1 = 2\sqrt{a}\cos\left(\frac{\Phi + n\pi}{3}\right); n = 0, \pm1, \pm2, \ldots$$

$$\Phi = arctg\left(\frac{\sqrt{a^3 - b^2}}{b}\right); a = \frac{1}{81}\left(1 + \frac{3(k^2 + 6)}{M^2}\right); b = \frac{1}{162}\left(\frac{2}{9} + \frac{k^2 - 12}{M^2}\right) \tag{13}$$

$$y = y_\pm = -\frac{y_1}{2} \pm \frac{y_2\sqrt{3}}{2};$$

$$y_2 = 2\sqrt{a}\sin\left(\frac{\Phi + n\pi}{3}\right) \tag{14}$$

Thus, taking into account (13) or (14), the exact real solution of the dispersion equation (11) has the form:

$$\omega_1 = \frac{4Mq_H(k;M)}{\pi}cl;$$

$$q_H = y(k;M) - \frac{8}{9} \tag{15}$$

The solution (15) must be considered together with the solution of the dispersion equation (9), which for the case $k > 0$ in the limit $lR_{s0} >> 1$ has the form:

$$\omega_1 + \frac{Mc}{R_{s0}} + \frac{(1+h)k^2c^2\delta}{2R_{s0}^2\omega_1} = \frac{(1 - 2M^2 - h)}{2M}cl \tag{16}$$

From (15) and (16) we obtain a solution in the form:

$$\omega_1 = \frac{4M^3q_H(k;M)c}{\pi R_{s0}(h_K - h)}\left(1 + \sqrt{1 + \frac{\pi^2(h_K - h)(1+h)k^2\delta}{4M^4q_H(k;M)}}\right); \tag{17}$$

$$h_K = 1 - 2M^2\left(1 + 4q_H(k;M)/\pi\right)$$

The condition for the realization of the corrugation instability $\omega_1 > 0; l > 0$ following from (16) and (17) has the form:

$$0 < q_H(k; M) = y(k; M) - \frac{8}{9} < \frac{\pi}{8M^2}\left(1 - 2M^2 - h\right),$$
$$h < 1 - 2M^2$$

(18)

In (18), it is necessary to use one of the three solutions (13), (14). For example, under conditions [*)]

$M = 0.562; M^2 = 0.316; h = -0.005$ corresponding to the observation data of a converging cylindrical shock wave obtained during an electric explosion of an ensemble of wires in water [12], the inequality (18) has the form:

$$0.888 < y(k) < 1.352$$

(19)

At the same time, in (13) and (14), it is necessary to use the representations:

$$\overset{..}{b} = 0.0195k^2 - 0.2330; a = 0.1172k^2 + 0.7156$$

(20)

From Fig. 2-Fig.4, where the dependence of functions (13) and (14) at conditions (20) are represented, it follows that only for sufficiently small numbers $0 \leq k \leq 0.25$ the condition (19) is valid and the instability of the shock wave front relative to two-dimensional perturbations for the parameter values under consideration is possible. For example, with the value $k = 0.25$ we get (see Fig.3) $y_1 \approx 1.351; q_H \approx 0.463; h_K \approx -0.00457; h_K - h \approx 0.00043$ and for the value $R_{s0} = 50 \times 10^{-6}$ m, we obtain from (17) the following estimate $\omega_1 \approx 54.5 \times 10^9 \sec^{-1}$.

For the value $k = 0.25$, the dependence of the perturbation amplitude on the angular variable is proportional to the function $\cos(0.25\varphi)$ that has a period $8\pi$ and at the same time the deviation of the front shape caused by such a perturbation is very small, but violation of cylindrical symmetry occurs. At the same time, it is important that only for $k \neq 0$ a small perturbation of the tangential component of the velocity field of the already linear stage of perturbation development increases exponentially rapidly in time with the specified increment. Therefore, in relatively short time intervals, this perturbation can reach a finite value and lead to a sufficiently intense rotation of the medium behind the shock wave front, which will prevent unlimited accumulation of the energy of a converging cylindrical shock wave [5].

As the value $k$ decreases, the value of the exponential index $\omega_1$ will decrease and at a zero value corresponding to one-dimensional perturbations, an estimate is obtained that is about 20 times less than the one obtained above (see the next paragraph).

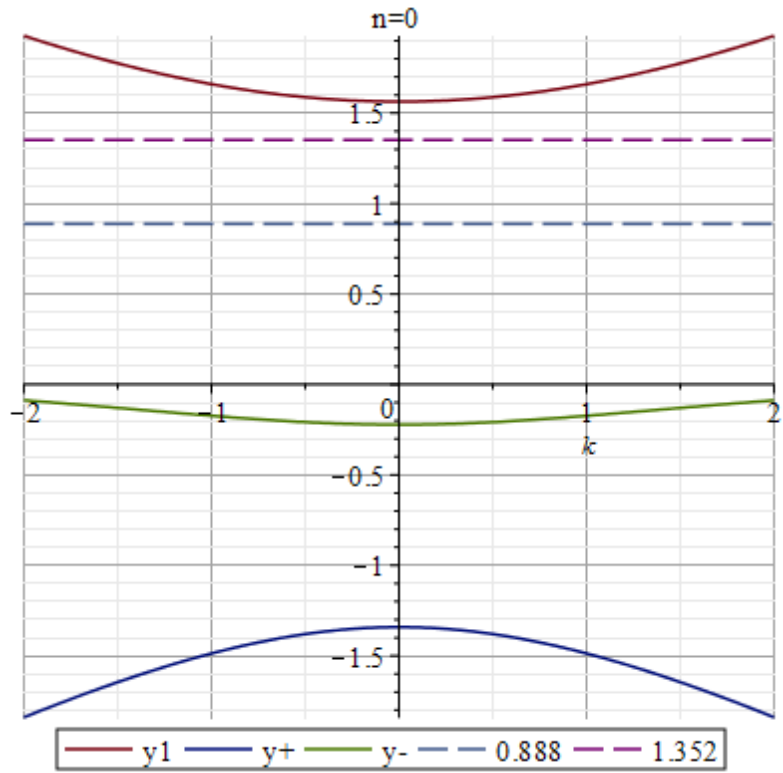

Fig.2.

Caption to Fig.2: Plots of functions $y_1$, $y_+$ and $y_-$ for $k \in [-2;2]$; n=0; M=0.562 according to (13), (14), (20). For this case instability is impossible.

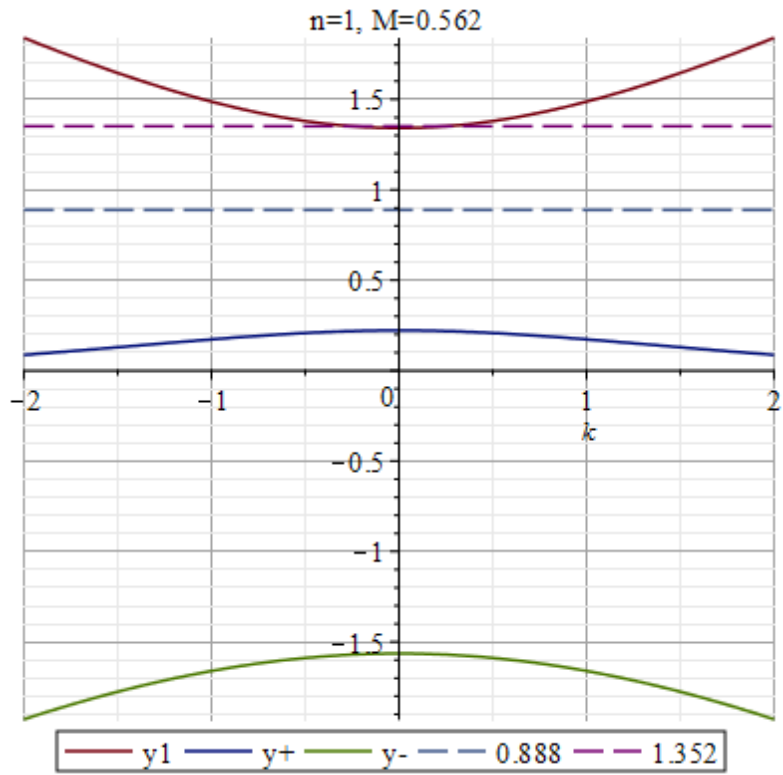

Fig. 3

Caption to Fig.3: Plots of functions $y_1(k)$, $y_+(k)$, and $y_-(k)$ for $k \in \left[-2; 2\right]$, n=1, M=0.562 according to (13), (14), (20). For this case instability is possible in finite range $0 \leq |k| \leq k_{ih} \approx 0.255$ (see below in Fig.4).

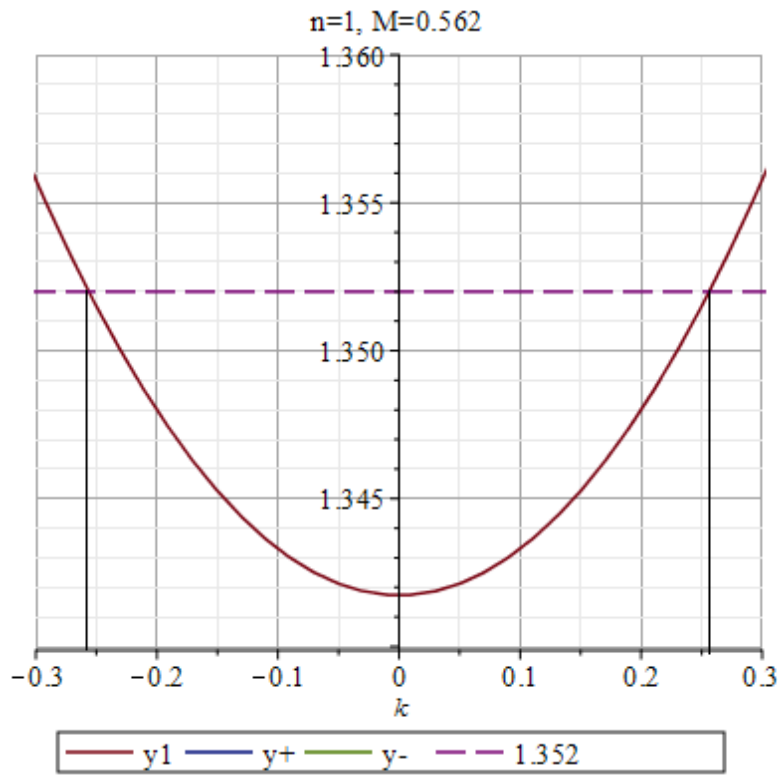

Fig.4

Caption to Fig.4: Here shows an enlarged fragment of the curve $y_1$ shown in Fig. 3.

In addition, we note only the qualitative correspondence of the conclusion obtained in this paper about the stability in the linear approximation for sufficiently large super threshold values ($k > 0.255$, see Fig.4) of the azimuthal wave number with the similar conclusion obtained by numerical analysis of the linear stability of the front of a converging spherical shock wave in [13]. Because in [13] the threshold spherical harmonic numbers is much larger than unity $l_c = 42.066$ for $\gamma = 5/3$ and $l = 80.295$ for $\gamma = 7/5$ above which eigenmode perturbations diminish in the converging process (see also Fig.6 in [13]).

## 2. One-dimensional instability

Let us consider separately the case of one-dimensional axially symmetric perturbations that do not violate the shape of the front of a converging cylindrical shock wave at all. It follows from the conditions of the experiment [12] that initially the perturbations have a two-dimensional character and depend on both the radial and angular variables (see Fig. 1). This is due to the finiteness of the number of exploding wires and the azimuthal inhomogeneity of the initial motion of the medium behind the shock wave front, which does not have an ideal cylindrical symmetry. However, after some time, such a cylindrical symmetry of the front of the converging shock wave is still established, as follows from the observational data. Therefore, when considering the question of the stability of the front of such a cylindrical shock wave, it is also of interest to take into account purely one-dimensional longitudinal perturbations that have azimuthal symmetry at a value $k = 0$ in (8)-(10) and do not violate the cylindrical symmetry. In this case, the tangential component of velocity field can also be nonzero, depending only on the radial coordinate according to (2) and (8). However, in this case, the tangential component of velocity is not related to the pressure perturbation and the perturbation of the radial velocity component in the linear approximation. Therefore, in contrast to the case of two-dimensional perturbations considered in the previous paragraph, for one-dimensional perturbations at the linear stage there is no increase in perturbations of the tangential component of the velocity field. Only at the nonlinear stage of perturbation development, this component of the velocity field becomes significant and determines the limitation of the internal energy accumulation value specified in the Introduction. In this connection, we consider the dispersion equations (9), (10), when one-dimensional (independent of the angular variable) extremely small perturbations do not violate the cylindrical symmetry of the shock wave front in the linear approximation. Indeed, it follows from the observational data that, as we approach the axis of symmetry, an azimuthally symmetric erosion of the observed shock wave front occurs in the radial direction with the preserved cylindrical symmetry of the front. This observed behavior of the front is most naturally described either with the help of one-dimensional perturbations with a zero azimuthal wavenumber $k = 0$, or with the help of two-dimensional perturbations with sufficiently small wavenumbers $k \leq 0.25$, as was done in the previous paragraph.

For zero azimuthal wave number $k = 0$ from the condition $\Omega^2 - 24c^2l^2 = 0$ in (10) and from (9) in the limit $lR_{s0} >> 1$, the index of exponential instability has representation:

$$\omega_1 = \frac{8cM^2}{R_{s0}\pi} \frac{\left(\sqrt{2/3} - M\right)}{\left(h_{th} - h\right)};$$

$$h_{th} = 1 + 2M^2(4/\pi - 1) - \frac{8M}{\pi}\sqrt{2/3}$$

(21)

$$l = \frac{2M^2}{R_{s0}(h_{th} - h)}$$

In (21) value $l^{-1}$ determines the characteristic distance at which there is an exponential decrease of the amplitude of perturbations in the radial direction according to (8).

The initial amplitude of the perturbation is small compared to the radius of the shock wave $R_{s0}$, which is assumed to be constant in the considered sufficiently small time intervals [15].

To obtain a solution in the form (21), it is necessary to take into account that for $k = 0$ the equality $\Omega^2 - 24c^2l^2 = 0$ is obtained from (10). This implies a representation $\omega_1 = 4cl\left(\sqrt{2/3} - M\right)/\pi$, and from (9) in the limit $lR_{s0} >> 1$ it is possible to obtain a solution in the form $\omega_1 = -Mc/R_{s0} + cl(1 - 2M^2 - h)/2M$. From these two relations for $\omega_1$ the solution in the form (21) is obtained.

In the limit $R_{s0} \to \infty$ in (21) is the lack of instability to a plane shock wave, as it should be for cases where the effects of viscous dissipation can be neglected [16], [18], [19].

Thus, from the inequalities $l > 0; \omega_1 > 0$ in (21) the instability conditions are obtained in the form:

$$M < M_{th1} = \sqrt{2/3};$$
$$h < -1/3 \qquad (22)$$

$$M < M_{th2} = a\left(1 - \sqrt{1 - b(1 - h)}\right); a = \frac{2\sqrt{2/3}}{4 - \pi}; b = \frac{3\pi(4 - \pi)}{16};$$
$$h > -1/3; \qquad (23)$$

For values $h < -1/3$ the inequality $M_{th1} < M_{th2}$ is realized and it is necessary to use the condition of instability (22). On the contrary, at values $h > -1/3$ the condition of instability of a converging cylindrical shock wave with respect to one-dimensional perturbations will already be fulfilled at $M_{th2} < M_{th1}$ and has the form (23).

For example, for the values of the parameters following from the simulation data (see *) after (18) and [12]) $M = 0.562; h = -0.005; D = 5100 m/\sec; c = 5600 m/\sec; \delta = 1.62; R_{s0} = 50 \times 10^{-6} m$, we obtain in (23) $M_{th2} \approx 0.568$, which correspond to the fulfillment of all the instability conditions in (23). At the same time, from (21) where $h_{th} = 0.004$ for the index of the exponential growth of perturbations, we obtain an estimate $\omega_1 \approx 2.49 \times 10^9 \sec^{-1}$.

The estimation of the compression time of the shock wave radius from radius $R_{s0} = 50 \times 10^{-6} m$ to radius $R_{s0} = 30 \times 10^{-6} m$ leads to the value $\Delta t \approx 3.92 \times 10^{-9} \sec$. As a result, during this time, the amplitude of any longitudinal (along the direction of compression) extremely small disturbance will grow by a factor of $e^{9.76} \approx 17327$. Note, however, that in the theory used here [15], only sufficiently small time intervals are considered, during which the radius of the shock wave can be considered as an approximately constant value. For example, such a time interval is equal $\Delta t \approx 0.98 \times 10^{-9} \sec$ when the shock wave radius changes from 50 to 45 microns. During this time, the amplitude of the initial perturbation will grow by $e^{2.44} \approx 12$ times. For the case of a change in the radius from 50 to 40 microns, the amplitude of the disturbances will increase by $e^{4.882} \approx 132$ times over time $\Delta t \approx 1.96 \times 10^{-9} \sec$.

Note that when taking into account the effects of viscous dissipation, the conditions for instability of the converging cylindrical wave front can be obtained according to the estimates given in [15] and the known observational data on the effect of compression on the value of the shear and volume (second) viscosity coefficients [22]-[24].

### 3. Discussion and comparison with known data.

The realization of instability significantly depends on the value of the Dyakov parameter $h$ and the Mach number $M$ as it follows from (18) and (22) or (23). However, in the experiment, it is possible to establish only the Mach number $M_0 = D/c_0$, determined by the magnitude of the velocity of the shock wave front relative to the undisturbed medium before the shock wave front. In this regard, to compare the conclusions of the theory about the conditions of exponentially rapid growth of perturbations of the tangential component of the velocity field leading to the rotation of the medium behind the shock wave front, it is necessary to establish a relationship between the magnitude of $M_0 = D/c_0$ and values $M = (D-U)/c$ and $h$ in conditions (18) and (22) or (23).

**Polytropic medium**

Let us consider, for simplicity, the case of one-dimensional perturbations, when the instability conditions (22), (23) are applied to a converging cylindrical shock wave in a polytropic medium characterized by an adiabatic index $\gamma$ [ [16]. In this case, the Dyakov parameter $h$, determined from (6) for a known type of the Hugoniot shock curve (see (89.1) in [16]), as well as the Mach number in the compression region have representations (see (89.9) in [16]):

$$h = -1/M_0^2$$

$$M^2 = \left((\gamma-1)M_0^2 + 2\right)\left(2\gamma M_0^2 - \gamma + 1\right)^{-1} \qquad (24)$$

At the same time, taking into account (24) instead of (22), we obtain the instability condition:

$$M_{0\min} = \sqrt{\frac{2(\gamma+2)}{\gamma+3}} < M_0 < \sqrt{3} \qquad (25)$$

Accordingly, from (23) and (24) follows the instability condition in the form:

$$F(M_0;\gamma) < 0; M_0 > \sqrt{3}$$
$$F = \sqrt{\frac{2+(\gamma-1)M_0^2}{2\gamma M_0^2 + 1 - \gamma}} - a\left(1 - \sqrt{1 - b\left(1 + \frac{1}{M_0^2}\right)}\right); \qquad (26)$$
$$a = \frac{2\sqrt{2/3}}{4-\pi} \approx 1.902352; b = \frac{3\pi(4-\pi)}{16} \approx 0.505644$$

These conditions define the one shown in Fig.5 the instability region, which is located above the curve 2 corresponding to the condition (25) and below the curve 1 determined from the instability condition (26).

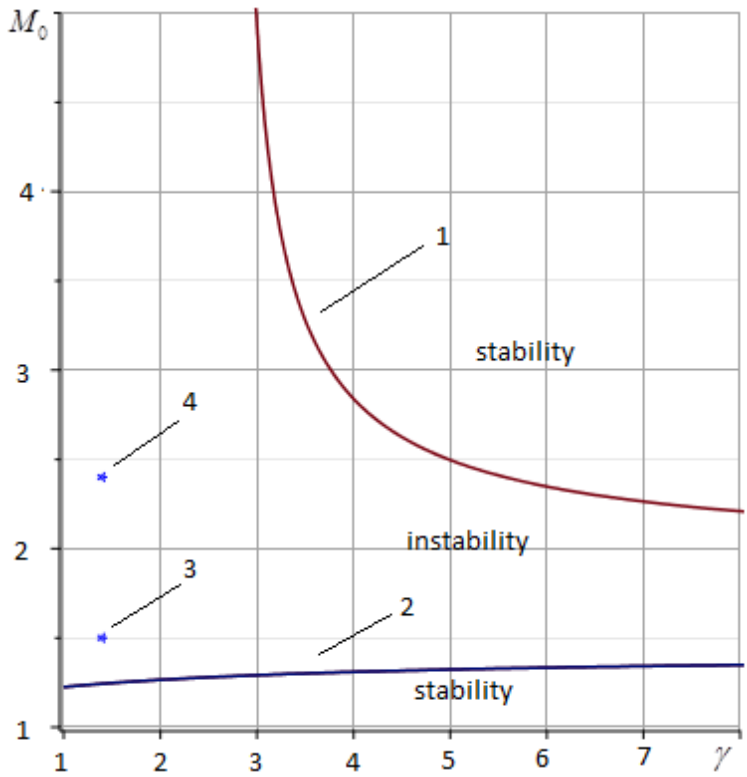

Fig. 5

Caption to Fig. 5,

the instability region is bounded from above by the curve of the dependence of the Mach number $M_0$ on the adiabatic exponent $\gamma$ (gamma), denoted by the number 1 and corresponding to the condition (26). The curve with the number 2 following from condition (25) restricts the instability region from below. The point with the number 3 corresponds to the data of the experiment and simulation [25] - [27] on the realization of the instability of a converging cylindrical shock wave in air. The point with the number 4 corresponds to the data of observations of the instability of a converging cylindrical shock wave in air [2]. Here, the value of the adiabatic index $\gamma = 1.4$ is used for air. In the limit $\gamma \to 1$ curve 2 limits to the value $M_{0\min} \to 1.225$ according (25).

At the same time, for the range of values of the adiabatic index $1 < \gamma < 2.7624$ for the realization of linear instability of a converging cylindrical shock wave in a polytropic medium, there is only a lower limit on the Mach number $M_0$ determined from condition (25). For this range of values $\gamma$, there is no longer a limit on the Mach number from above, since curve 1 has an asymptote $\gamma \approx 2.7624$ corresponding to the limit $M_0 \to \infty$.

Above, restrictions on the Mach number were obtained, which follow from the instability conditions with respect to one-dimensional perturbations (22) or (23).

Note that for a fixed value of the azimuthal wave number $k > 0$, from the left inequality (18), at which the instability of a converging cylindrical shock wave can be realized the generalization for condition (22) is obtained. For example, for the threshold case $k = 0.25$ considered above, we obtain the following necessary instability condition in the form $M < M_{th}(k = 0.25) \approx 0.8228$.

It can be seen from this that the threshold value $M_{th}(k = 0.25)$ does not differ much from the threshold value $M_{th1} = \sqrt{2/3} \approx 0.8165$ in (22), which determines the threshold of instability with respect to one-dimensional perturbations at the value $k = 0$. At the same time, the corresponding lower threshold value of the Mach number for two-dimensional corrugation perturbations differs little, for which, taking into account (24) , we obtain a generalization of condition (25) in the form:

$$M_0 > M_{0th} = \left( \frac{2 + M_{th}^2(k)(\gamma - 1)}{2\gamma M_{th}^2(k) + 1 - \gamma} \right)^{1/2} \qquad (27)$$

In the case $M_{th}(k = 0) = M_{th1} = \sqrt{2/3}$ the condition of instability (27) coincides with (25).

For two-dimensional disturbances from inequality (18) it is also possible to obtain the generalization of (23).

For the above estimate from (27) for the value $M_{th}(k = 0.25) \approx 0.8228; M_{th}^2 \approx 0.677$, we obtain that $M_{0th} \approx 1.32$ for $\gamma = 7$ .According to (27), with the same value of the wave number, the threshold value of the Mach number for the case of a gas with an adiabatic index $\gamma = 7/5$ ( (for example, for air) has a value $M_{0th} \approx 1.23$ that differs little from the value $M_{0\min} \approx 1.243$ obtained from (25) for one-dimensional perturbationsd value $M_{0th} \approx 2.4$ (see point 4 in Fig.5) noted in [2] and the theoretical estimation. Note that these estimates made on the basis of (25) and (27) are qualitatively and quantitatively consistent with the instability effect of a converging cylindrical shock wave in air observed at Mach numbers near $M_0 \approx 1.5$ [25]-[27] (see point 3 in Fig. 5). The boundary for the threshold Mach number which is obtained from (25) and (26) in Fig.5 is also qualitatively in agreement with the threshold Mach estimation $M_{0th} \approx 2.531$ in [1]. In [1] for the plane shock in air travelling along a wall with convex corner the point of inflection is arising on its front when the Mach number has super threshold value $M_0 \geq 2.531$ (see after (55) in [1]).

**Arbitrary medium**

In the case of an arbitrary medium, to determine the Dyakov parameter, it is possible to use the well-known representation for the Hugoniot shock curve, which follows from the experimental data on the linear relationship between the velocity of the shock wave and the velocity of the medium behind the shock wave front, which has the form [28]-[33]:

$$D = A + BU \qquad (28)$$

For example, for water, in the speed range $7.1 > U > 1.5$ km/s in (28) $A = 2.393$ km/s, $B = 1.333$, [31], and for the range $0.3875 > U > 0.068$ km/s, it is already necessary to use other values $A = 1.45$ km/s, $B = 1.99$ km/s, [32].

Taking into account (28) and on the basis of the known Rankin-Hugoniot relations

$p/p_0 = 1 - DU\rho_0/p_0; 1/\delta \equiv \rho_0/\rho = 1 - U/D$ on the shock [16], from the definition (6), we obtain for the Dyakov parameter the ratio:

$$h = -\frac{A}{c_0\left(2M_0 - \dfrac{A}{c_0}\right)} \qquad (29)$$

Using (28), we also obtain the relations for the compression value $\delta = \rho/\rho_0$ and the Mach number $M$ in the compression region:

$$\delta = \frac{M_0 B}{M_0(B-1) + A/c_0} \tag{30}$$

$$M = \frac{M_0(B-1) + A/c_0}{Bc/c_0} \tag{31}$$

In addition, to determine the speed of sound in the compression region, we will use the ratio $c^2 = c_0^2 \delta^{n-1}$ that corresponds to the well-known isentropic equation of the state of water $p - p_0 = \frac{\rho_0 c_0^2}{n}\left(\delta^n - 1\right)$ [34], [35]. In this case, we obtain the following representation:

$$M^2 = \left(M_0(1 - \frac{1}{B}) + \frac{A}{c_0 B}\right)^{n+1} M_0^{-(n-1)} \tag{32}$$

For example, from the instability conditions $M^2 < 2/3$ and $h < -1/3$ corresponding to (22), from (29) and (32) under the conditions $n = 7$, $A = 2393m/\sec; B = 1.333; c_0 = 1500m/\sec$, we obtain the instability condition in the form $2A/c_0 \approx 3.19 > M_0 > M_{0th} \approx 2.3$.

Note that in addition to the lower threshold Mach number from (22), there follows the existence also of an upper threshold Mach number $M_{0th}^{up}$ from (23) for the case $h > -1/3$ when (29) gives $M_0 > 2A/c_0 \approx 3.19$ and instability is possible only under the condition $3.19 < M_0 < M_{0th}^{up}$. For the case considered above, this restriction is mainly of a formal nature, since at the specified parameter values $A; B$, this threshold value is equal to a very large value $M_{0th}^{up} \approx 145$.

Note, however, that when considering the relation (23) between the velocity of the shock wave $D$ and the velocity of the medium behind the shock wave front $U$ in a narrower range of medium velocities, the value of the parameters $A; B$ can take other values. For example, in the range $1588m/\sec \le U \le 2108m/\sec$ according to the data in the water [33] (see Table III in [33]) with an accuracy of at least 6%, in (23), the values $A \approx 50.371m/\sec$ and $B \approx 2.587$, for which the Dyakov parameter already takes a value $h \approx -0.005$ at the compression value $\delta = 1.62$ can be used. In this case according to (23) and (32) instability arising only for Mach numbers bounded from above $M_0 \le M_{0th}^{up} \approx 3.92$. This upper threshold value for Mach numbers significantly exceeds the upper limit ,

$M_0 \le M_{0th}^{up}(\gamma = 7) \approx 2.25$, which is obtained according to Fig.5, if we consider water in the approximation of a

polytropic medium with an adiabatic $\gamma = n = 7$.

Indeed, qualitatively, such existence of restrictions on the Mach number from above and from below is in accordance with result showing in Fig.5 for $\gamma > 2.8$ and the data of experimental observations of the instability of plane shock waves in gases, given in [36] and in theory for plane shock waves in water [21] (see (6.5) in [21]).

At the same time, the conclusion obtained here about the existence of an upper threshold Mach number for the possibility of realizing linear instability is in qualitative accordance with the dependence of the stability region on the Mach number obtained in [6]. Indeed, in [6], a conclusion was obtained about the boundary to stability region expansion of spherical converging waves in argon with an increase in the Mach numbers. For example, according to [6] (see Fig. 12 in [6]), for the Mach number $M_0 = 4.0$, the maximum permissible for stability relative amplitude of disturbances along the radius of the front of a converging spherical shock wave in an ideal polytropic gas is equal to $\Delta r / R_{s0} \approx 0.25$. So, at this Mach number the instability of the front already occurs for the perturbation larger amplitude $\Delta r / R_{s0} = 0.3$. However, for the larger Mach number $M_0 = 8.0$ with the same perturbation amplitude and for the same perturbation form, the stability of the front is preserved [6]. For a converging cylindrical shock wave in water, the front shape is also preserved at Mach numbers near $M_0 \approx 6$ in experiment [37].

It should be clarified here that in [6] only the stability of the shape of the shock wave front is considered. This,

however, does not exclude the possibility of forming a rotation mode of the medium behind the shock wave front,

which, without affecting the shape of the front, can limit the accumulation of internal energy near the axis of convergence of the shock wave according to the mechanism considered in [5] or under conditions (18) and (22),

(23) exponentially rapid growth of small perturbations (see also [38]).

**Cumulation restriction for a converging cylindrical shock wave in water**

Let us consider the application of the results obtained in this paper to interpret the conclusion [12] about the possibility of a noticeable restriction of pressure accumulation for a converging cylindrical shock wave in the case of a sufficiently large azimuthal wave length $\lambda$ of initial disturbances during an underwater electric explosion of wire ensembles. Indeed, according to [12] (see Fig. 7a in [12]), a noticeable limitation of cumulation is realized only in the case of the largest-scale perturbations of the dipole type, for which the azimuthal dimensionless wave number is equal to $k = \dfrac{2\pi R_{s0}}{\lambda} = 2$. The perturbation wavelength is equal to $\lambda = 2\pi R_{s0}/k \approx 0.314 \times 10^{-3} m$ when $k = 2$ and $R_{s0} = 0.1 \times 10^{-3} m$ (see Fig.6b in [12]). Note that the radius of the wire ensemble before the explosion is equal to $R_E = 2.37 \times 10^{-3} m$ [12], when $\lambda_{\max} \approx 7.4 \times 10^{-3} m$ for the case $R_{s0} \approx R_E$.

The specified dipole character of the artificial disturbance of the shock wave front is obtained by choosing the appropriate power distribution allocated in each of the four segments of the wire ensemble consisting of only forty thin copper wires, when energy is released in each segment that is different from the energy released in the neighboring segment. At the same time, approximately the same energy is released in oppositely located segments during an underwater explosion of wires. As a result, macro non-uniformities can be developed relatively far from the converging axis because of different (up to 40%) rates of heating of the wires, resulting in their non-simultaneous explosion [12] (see Fig.6b in [12]).

In the case of a fixed wave number, the necessary instability condition that determines the threshold Mach number follows from the left side of inequality (18), as noted in the previous paragraph. In particular, for the case of one-dimensional perturbations with a wave number $k = 0$ for the shock wave radius $R_{s0} = 50 \times 10^{-6} m$, the shock wave velocity according to the data [12], [39]-[41] is equal to $D = 5100 m/\sec$, which corresponds to the Mach number $M_{0th}^{up} \approx 3.92 > M_0 \approx 3.4 > M_{0th} = 2.3$ and the fulfillment of the necessary condition for exponentially rapidly developing instability according to (22) and (32).

Note that for two-dimensional corrugation perturbations, the specified Mach number in [12] also satisfies the necessary condition of exponential instability for the case when the wave number is equal to $k = 0.25$, which is close to the maximum allowable value for a fixed Mach number (see Fig.4). This value of the azimuthal wavenumber for the considered shock wave radius $R_{s0} = 50 \times 10^{-6} m$ corresponds to the wavelength $\lambda \approx 1.26 \times 10^{-3} m$, which is commensurate with the above-mentioned wavelength range (from $\lambda \approx 0.314 \times 10^{-3} m$ to $\lambda_{max} \approx 7.4 \times 10^{-3} m$) for the dipole-type perturbation in [12] (see Fig.6b in [12]). This correspondence between the wavelengths may indicate the existence of a mechanism by which, during the compression process, the wavelength of the perturbation remains a little-changing, almost invariant value. This issue is of independent interest and requires separate consideration.

As a result, we can give the following interpretation of the noted conclusion of [12] on the limitation of cumulation. According to our results the instability limiting the cumulation is associated precisely with the fulfillment of conditions under which an exponentially rapid growth of perturbations occurs even under the assumption about of an almost unchanged shock wave radius. This interpretation differs significantly from the interpretation given in [12], according to which even with a weak power - law growth of perturbations, it takes quite a long time for their development when considering perturbations with the largest possible wavelength-perturbations of the dipole type. At the same time, it is essential that the initial perturbations with large wavelengths (perturbations of the dipole type in [12]) retain information about themselves under strong compressions, where even the extremely small amplitude of such perturbations can grow exponentially quickly to finite values in a very short time and lead to a limitation of cumulation. It is possible that even a weak power-law growth of these large-scale perturbations allows them to survive until at the sufficient strong compression, the conditions (18) and (22) or (23) for their exponentially rapid growth begin to be met, which leading to an intensification of the rotation of the medium behind the front and a corresponding limitation of cumulation.

At the nonlinear stage of perturbation development, it is possible to form a rotation mode of the medium behind the front of a converging cylindrical shock wave. The energy of radial compression is transformed into it, which limits the energy accumulation during implosion. For experimental verification of this conclusion, it is necessary to develop methods for directly measuring the direction and magnitude of the velocity of the medium behind the shock wave front.

In this regard, a new interpretation can be given for the already known data of experimental observations and simulations. At the same time, it should be borne in mind that the resulting rotational motion of the medium behind the shock wave front at sufficiently high rotational speeds can itself become unstable and the observations will no longer record the initial rotation of the medium itself, but the result of such a secondary instability of this rotation. For example, it can be an instability that develops according to the type of Taylor-Couette instability, which occurs at an over-threshold speed of rotation of a liquid between two coaxial cylinders rotating at different speeds [16], [42]-[44]. As is known [42], as the rotation frequency increases, instability manifests itself in the sequential appearance of one, two and three vortex modes, the interaction of which, with a subsequent increase in the rotation frequency, leads to developed turbulence characterized by a continuous energy spectrum. Indeed, during experimental observations and simulations for a converging cylindrical shock wave in air, it was found that during implosion, systems of three and four vortex structures appear in the medium density field behind the shock wave front and, accordingly, distort the initial symmetry of the front when it approaches the axis of symmetry [25]-[27]. At the same time, the method of holographic interferometry was used to quantify variations in the density of the medium in the compression region [25]. However, other observation methods also lead to the establishment of a similar four-vortex structure near the convergence axis of a cylindrical shock wave in a gas [45] (see Fig. 4.3 a and Fig. 4.4 in [45]). Note that for a converging cylindrical shock wave in water during implosion, a front structure similar to a single spiral vortex mode is observed [41] (see Fig. 3e in [41]), which violates the initial symmetry of the front to a much greater extent than it can be due to the technical optical effect mentioned in [41] in this regard. Indeed, for the values of 125 microns and 75 microns characterizing the lengths of the semi-axes of the ellipsoidal structure shown in Fig.3 e in [41], there is a much greater difference than the distortion value of about 7 microns, which corresponds to the discrepancy value $(10^{-2})^0$ indicated in [41] between the assembly axis and the optical axis of the recording camera. At the same time, judging by the direction of the spiral arms in Fig.3 e, the direction of rotation of the vortex structure represented there corresponds to a clockwise rotation.

In this connection, we will assume an analogy between the rotation of a layer having a thickness $h_R$, behind the front of a converging cylindrical shock wave of radius $R$, with the Taylor-Couette flow between an inner cylinder rotating with a frequency $\Omega$ having a radius $R$ and a stationary outer cylinder having a radius $R + h_R$. For this case, the instability of the Taylor-Couette flow occurs under the condition [16] (see (27.5) in [16]):

$$\mathrm{Re} = \frac{h\Omega R}{\nu} > \mathrm{Re}_{th} = 41.3 \sqrt{\frac{R}{h}} \qquad (33)$$

It is possible to use (33) for obtain the condition of instability of the rotation of the medium behind the shock wave front. Let us assume that at the radius $R = 50 \times 10^{-6} m$ of the shock wave, the velocity of the radial motion of the medium behind the shock wave front $U = 1950 m/s$ is equal in magnitude to the velocity of rotation of the medium when the relation $\Omega R \approx U$ is fulfilled in (33). In (33), we use an estimate $\nu \approx 10^{-3} m^2/s$ (see [22]-[24]) for the kinematic viscosity coefficient for water behind the shock wave front that exceeds its value under normal conditions (when $\nu \approx 10^{-6} m^2/s$). In this case, according to (33), even for such an overestimated value of the kinematic viscosity coefficient, the instability of the rotation of the medium behind the front will occur under the condition $h_R > h_{Rth} \approx 1.31 \times 10^{-6} m$.

Thus, when the rotation of the medium occurs behind the shock wave front, this mode of motion of the medium is obviously unstable, which can lead to the observed in [41] and [25], [45], [46] vortex structures. At sufficiently large initial Mach numbers, these vortex structures, in turn, become unstable, which leads to the stochastic turbulent regime, similar to that observed for the Taylor-Couette flow [42].

## Conclusions

The conditions for exponentially fast growth of extremely small perturbations are obtained for converging cylindrical shock waves in arbitrary media and under the broadest assumptions about the nature of the undisturbed motion of the medium in the compression region behind the shock wave front. This is the main difference between the proposed linear theory and the usual consideration of only an ideal polytropic gas and a self-similar solution of the second kind for describing the undisturbed motion of the medium behind the front. To date, such a generality of the shock wave stability theory has been realized only in the Dyakov theory [16], [18], [19] for the case of plane shock waves and in [15] for the cylindrical converging shock waves.

The obtained linear instability of a converging cylindrical shock wave weakly violates the cylindrical symmetry of the shape of the shock wave front. As a result of the corresponding rapid exponential growth of small perturbations of the tangential component of the velocity of the medium behind the front, it is possible to form a rotation of the medium, which leads to a restriction of the accumulation of internal energy during implosion. The comparing of the developed theory with the observational data of converging cylindrical shock waves in water is also provided and gives interpretation of that data. In this regard, it is of interest to develop in future a theory that gives an analytical description of the equation of state of water, by using virial theorem [38], as well as to compare in more detail with the data of experimental observations of shock waves that occur during underwater explosions.

I thank Ya. E. Krasik for the permanent support and attention to work. I also express my gratitude to A. Virozub for providing simulation data and to A. G. Chefranov for help in obtaining of Fig. 1-Fig.5.

The study is supported by the Israel Science Foundation, Grant number: 492/18

Conflict of interest: "The authors have no conflicts to disclose".

Data Availability Statements: "The data that support the findings of this study are available from the corresponding author upon reasonable request.